\documentclass[%
 aip,
 amsmath,amssymb,
 reprint,
 onecolumn,
 pof
]{revtex4-1}

\usepackage{graphicx}% Include figure files
\usepackage{dcolumn}% Align table columns on decimal point
\usepackage{bm}
\usepackage{amsmath}
\usepackage[english]{babel}
\usepackage{amsfonts} 
\usepackage{xcolor}
\usepackage{dsfont}
\usepackage[normalem]{ulem}
\usepackage[utf8]{inputenc}
\usepackage[T1]{fontenc}
\usepackage{mathptmx}
\usepackage{etoolbox}
\UseRawInputEncoding
\usepackage{graphicx}
\usepackage{tikz}
\usepackage{amsmath}
\usetikzlibrary{quantikz}

\makeatletter
\def\@email#1#2{%
 \endgroup
 \patchcmd{\titleblock@produce}
  {\frontmatter@RRAPformat}
  {\frontmatter@RRAPformat{\produce@RRAP{*#1\href{mailto:#2}{#2}}}\frontmatter@RRAPformat}
  {}{}
}%
\makeatother

\begin{document}

\preprint{AIP/123-QED}

\title[Carleman-lattice-Boltzmann quantum circuit with matrix access oracles]{Carleman-lattice-Boltzmann quantum circuit with matrix access oracles}
% Force line breaks with \\
\author{Claudio Sanavio}
\affiliation{Fondazione Istituto Italiano di Tecnologia\\
Center for Life Nano-Neuroscience at la Sapienza\\
Viale Regina Elena 291, 00161 Roma, Italy}
 \email{claudio.sanavio@iit.it}
%\author{Alexis Ralli}
%\affiliation{Department of Physics and Astronomy, Tufts University\\
%574 Boston Avenue, Medford, MA 02155, USA}
\author{William A. Simon}
\affiliation{Department of Physics and Astronomy, Tufts University\\
574 Boston Avenue, Medford, MA 02155, USA}
\author{Alex Ralli}
\affiliation{Department of Physics and Astronomy, Tufts University\\
574 Boston Avenue, Medford, MA 02155, USA}
\author{Peter Love}
\affiliation{Department of Physics and Astronomy, Tufts University\\
574 Boston Avenue, Medford, MA 02155, USA}
\author{Sauro Succi}
\affiliation{Fondazione Istituto Italiano di Tecnologia\\
Center for Life Nano-Neuroscience at la Sapienza\\
Viale Regina Elena 291, 00161 Roma, Italy}
\affiliation{Department of Physics and Astronomy, Tufts University\\
574 Boston Avenue, Medford, MA 02155, USA}

\date{\today}% It is always \today, today,

\begin{abstract}

We apply Carleman linearization of the Lattice Boltzmann (CLB) representation of fluid flows 
to quantum emulate the dynamics of a 2D Kolmogorov-like flow.
We assess the accuracy of the result and find a relative error of the order of $10^{-3}$ with just two
Carleman iterates, for a range of the Reynolds number up to a few hundreds.
We first define a gate-based quantum circuit for the implementation of the CLB method and then 
exploit the sparse nature of the CLB matrix to build a quantum circuit based on 
block-encoding techniques which makes use of matrix oracles. 
It is shown that the gate complexity of the algorithm is thereby dramatically 
reduced, from exponential to quadratic. 
However, due to the need of employing up to seven ancilla qubits, the
probability of success of the corresponding circuit for a single time step 
is too low to enable multi-step time evolution. 
Several possible directions to circumvent this problem are briefly outlined.
 
\end{abstract}

\maketitle

\section{Introduction}\label{sec:I}

The Lattice Boltzmann method (LBM) has proved to be an important tool for the simulation of 
fluid dynamics with a wide range of applications in the physics of fluids and allied disciplines \cite{benzi_lattice_1992,succi_lattice_2018,mendoza_fast_2010,aidun_lattice-boltzmann_2010,tran_lattice_2022,dunweg_lattice_2009}. 

The basic equations of classical fluid motion are known for over two centuries,
after the seminal work of Louis Navier in France and Gabriel Stokes in UK.
In essence, they are Newton's equation in reverse, $ma=F$, as applied to
a finite volume of fluid.
For the case of an incompressible flow (constant density), they take 
the following form:
\begin{equation}
\label{eq:NSE}
\partial_t \vec{u} + \vec{u} \cdot \nabla \vec{u} = -\nabla p + \nu \Delta \vec{u}  
\end{equation}
where $\vec{u} = \vec{u} (\vec{x},t)$ is the space-time dependent flow field (vector),
$p$ is the pressure and $\nu$ the kinematic viscosity.
Incompressibility forces the solenoidal condition
\begin{equation}
\label{eq:CONTI}
\nabla \cdot \vec{u} = 0. 
\end{equation}

Despite their innocent appearance, the Navier-Stokes equations (NSE) hide a 
Pandora's box of complexity, from the dripping
droplets in the kitchen faucet, all the way up to geophysical flows, including 
extreme events such as tornadoes and hurricanes, a very hot topic 
in these modern times of environmental concerns. 

The key mechanism is deceivingly simple: the nonlinear and nonlocal term 
$\vec{u} \cdot \nabla \vec{u}$ fuels a cascade of energy 
from large to small scales, virtually dissipation-free.
Dissipation enters stage only once the fluid structures become small enough for dissipation
to take over the nonlinear energy transfer. The crossover scale, named after the russian 
polymath Andreij Kolmogorov is given by $L_k = L/Re^{3/4}$ where 
\begin{equation}\label{eq:Reynolds}
Re=\frac{UL}{\nu}
\end{equation}
is the so-called Reynolds number, the effective measure of 
Nonlinearity versus Dissipation. In the above, $L$ is the macroscopic scale of the
fluid, $U$ its typical macroscopic velocity and $\nu$ the viscosity of the fluid.

The Reynolds number is generally pretty large; an ordinary car 
features $Re \sim 10^7$ and that's where complexity steps in: the number of active 
structures in a flow at Reynolds 
$Re$ scales like $(L/L_k)^3 \sim Re^{9/4}$ and computing such flow over a time span
$T = O(L)$ involves $O(Re^3)$ operations, the computational complexity of fluid turbulence.

For a car this means circa $10^{24}$ floating-point operations per simulation
(we count about 1000 operations each grid site and time step), a run that a perfect
Exaflop computer ($10^{18}$ Flops/second) would complete in about two weeks. 
Hence we may feel confident than in a  few years, classical Exascale computers
may "compute away" a fully-fledged car design.
Not so for weather forecasting, where Reynolds numbers increase 
by another three to five orders of magnitude.
Whence the ceaseless race to new and better computational methods.
Including, lately quantum computing\cite{nielsen_quantum_2010}.

Quantum computers may offer a solution to the complexity problem, since the dimension of Hilbert's 
space scales exponentially with the number of qubits.  
In his 1982 paper "Simulating physics with computers" \cite{feynman_simulating_1982}, Richard 
Feynman argued that "nature isn't classical", hence if we want to simulate nature
at a fundamental level we'd better use quantum computers. 
%Hence, he proposed to use quantum computers to simulate quantum dynamics, but he 
%did not explicitly advocate the use of quantum computers to simulate classical physics alike. 

Feynman invoked the QQ approach to scientific computing: quantum computers for
quantum physics. Fluid turbulence however is (mostly) classical, hence a CC
approach (classical computers for classical systems) appears reasonable
after all. Yet, given the large Reynolds numbers encountered in science and engineering, the 
computational complexity $Re^3$ is steep, hence it is of decided interest
to investigate whether a QC approach (quantum computers for classical physics) may 
offer a quantum advantage. 

In principle the potential is huge: since qubits offer logarithmic scaling for the 
$Re^{9/4}$ complexity, $\frac{9}{4} log_2(Re)$ qubits
are in principle sufficient to represent a given Reynolds number $Re$.
This would be mind-boggling with potentially revolutionary perspectives: 
with 10 to 100 physical qubits for each logical one, local weather forecast ($Re \sim 10^{10}$)
would become available with less than thousands to ten thousands qubits.

However, realizing this blue sky scenario faces a sequence of steep challenges:
first, one must devise a quantum algorithm for fluids, second it must show logarithmic
scaling of its computational complexity, third it must run efficiently on actual quantum
hardware. In the following, we take a very optimistic stance, namely that efficient 
quantum hardware will eventually become available, far from foregone
given the levels of quantum noise in current day quantum hardware.
Hence, in the following, we focus on the task of devising a quantum computing 
algorithm for the fluid equations.  

The task of simulating classical fluid dynamics on quantum computers requires the ability
to handle nonlinearity and dissipation. 
Quantum computers behave according to the laws of quantum mechanics, which are both linear and non-dissipative. 
Many strategies have been proposed in the recent years to handle both items above~\cite{succi_quantum_2023}, but it appears
fair to say that, to date, none of them has led to a viable quantum algorithm for fluids, meaning by this an
algorithm that can be run efficiently on actual quantum hardware.

This paper follows in the wake of a series of works on various approaches to implement a quantum Lattice Boltzmann algorithm~\cite{itani_quantum_2024,sanavio_lattice_2024}, and on the Carleman linearization procedure~\cite{itani_analysis_2022,sanavio_three_2024,sanavio_carleman-grad_2024}. 
In our previous work, we found that the Carleman linearization (CL) is able to approximate nonlinear fluid dynamics 
with an error that decays exponentially with the number of Carleman iterations~\cite{sanavio_explicit_2024}, a result consistent with the findings in Refs. ~\cite{liu_efficient_2021,liu_efficient_2023}. 
In particular, when applied to the Lattice Boltzmann model, the Carleman procedure shows 
an excellent convergence: with a second truncation level, CL represents the exact dynamics 
of 1D model~\cite{itani_analysis_2022}, and a relative error in the order of $10^{-3}$ for a 2D Kolmogorov-like flow~\cite{sanavio_lattice_2024,sanavio_three_2024}. 
With a truncation level at third order, CL exhibits a relative error of the same of order 
of magnitude as the intrinsic error of Lattice Boltzmann~\cite{li_potential_2023}. 

These favourable properties motivated the search of a Carleman Lattice Boltzmann (CLB) quantum algorithm.
In~\cite{sanavio_lattice_2024}, an algorithm is developed based on linear combination of 
unitaries (LCU) to implement the Carleman evolution. 
For a single timestep evolution, the gate complexity of the system was fixed to 
few thousands CNOTs, regardless of the number of lattice sites. 
However, the implementation of a multiple timestep simulation requires to measure 
and reinitialize the circuit at each timestep, a process that would require an exponential number of measurements. 
Conversely, the multi-time step circuit led to a gate complexity 
that scales exponentially with the number of qubits. 
In~\cite{li_potential_2023}, the authors calculated the gate complexity of a 
Carleman Lattice Boltzmann algorithm and showed that it can be made linear in time and logarithmic 
with the number of qubits, under the assumption that one is able to prepare 
an efficient oracle that embeds the Carleman evolution. 

In this paper, we describe an efficient preparation of the circuit that implements 
the multiple-time step Carleman Lattice Boltzmann evolution. Our approach is based upon a
block-encoding technique, and we show it is a potential way out of the problem, since 
it reduces the gate complexity from exponential to polynomial.

In Sec.~\ref{sec:II}, we describe the CL procedure that we use to 
transform the nonlinear problem into a linear one. 
In Sec.~\ref{sec:III} we apply CL to the LB for the case of a 2D Kolmogorov-like flow and 
we discuss the convergence of the CLB algorithm. In Sec.~\ref{sec:IV} we discuss the potential obstacles to an implementation of the CLB algorithm. 
In Sec.~\ref{sec:V} we describe the block-encoding technique as applied to the  
CLB framework and provide the explicit form of the quantum circuit, along 
with an analysis of its performance. 
In Sec.~\ref{sec:VI} we draw preliminary conclusions and portray a few promising
directions for future developments. 

\section{Carleman linearization}\label{sec:II}

The basic idea of Carleman linearization is to embed a finite-dimensional 
non-linear problem into an infinite-dimensional linear one.
The idea is readily explained by means of the zero-dimensional
logistic equation:

\begin{eqnarray}
\label{eq:LOGI}
\dot u = -u(1-Ru),\\ 
u(0)=u_0
\end{eqnarray}
where $u$ is a single variable (zero spatial dimensions) and $R$
measures the strength of the nonlinearity.

The Carleman embedding consists in renaming $u_1 \equiv u$ and $u_2 \equiv u^2$,
so that the logistic equation turns into $\dot u_1 = -u_1 + Ru_2$, which
is linear but open, since $u_2$ is a new independent variable.
The equation for $u_2$ is easily found by multiplying Eq.~\eqref{eq:LOGI} by $u$, to
obtain $\dot u_2 = -2(u_2 -Ru_3)$ where $u_3 \equiv u^3$. 
The trick is now clear, at each step $k$ one generates an open equation for $u_k$
involving $u_{k+1}$, thus leading to an infinite Carleman chain.
Eventually, this infinite chain can be solved analytically  
to recover the exact solution, but in general one looks for
a finite-order truncation, hopefully providing a reasonably accurate approximation.
By truncating at a given order $K$, one obtains:
\begin{eqnarray}
\dot u_k &=& -k (u_k-Ru_{k+1}),\;\quad k=1,\dots,K-1\\
\dot u_K &=& -K u_K,
\end{eqnarray}
with initial conditions $u_k(0)=u_0^k$.

%comment: This is true only when R is negative, which is not in the case of fluid dynamics, where the system tends to relaxation as shown in Carleman-Grad paper.

The bet is to achieve reasonable accuracy within a small number of iterations
along the Carleman ladder. For a thorough analysis of the Carleman linearization applied to the logistic equation we refer to~\cite{sanavio_carleman-grad_2024}.

The idea is elegant but faces a number of issues when applied
to multi-dimensional problems.
To appreciate the point, let us consider the Navier-Stokes equations, after spatial 
differencing and Euler time-marching:

\begin{equation}
u(t+h) = (I + Lh) u(t) + h Re \; Q \;u(t)u(t) \equiv A u + B u\otimes u
\end{equation}
where $u$ is the state vector consisting of $N=O(G)$ variables 
in a discrete grid of $G$ lattice sites. In the above, $I$ is the identity, 
$L$ is the discretized Laplacian, $Re$ is the Reynolds number and $Q$ is the
discretized quadratic advection operator.

For notational simplicity we have set $A=(I+hL)$ and $B = h Re Q$, where
$A \equiv A_{ij}$ is a rank-2 sparse matrix and $B \equiv B_{ijk}$ 
is also a sparse matrix of rank 3.
For simplicity, we have left out pressure, hence the above scheme applies 
to the Burgers equation.
 
Call $U_2\equiv u\otimes u$ and similarly $U_3,U_4,\dots$, the Burgers 
becomes a linear, open equation.
\begin{equation}
U_1(t+h) = A U_1(t) + B U_2(t) .
\end{equation}
The equation for $U_2$ is readily obtained:
\begin{equation}
U_2(t+h) = A^{\otimes 2} U_2(t) + Re (A\otimes B+B\otimes A) U_3(t) + B^{\otimes 2} U_4(t).
\end{equation}

The above set of equations can be written in terms of a linear system 
\begin{equation}
\begin{pmatrix}
U_1\\
U_2\\
\vdots
\end{pmatrix}(t+h) = \mathcal{C}\begin{pmatrix}
U_1\\
U_2\\
\vdots
\end{pmatrix}(t),
\end{equation}

\noindent where $\mathcal{C}$ is the Carleman matrix.
The lowest possible truncation ($K=2$) of the above system
leaves us with the following system:
\begin{eqnarray}
\label{C2}
U_1(t+h) &=& AU_1(t) + B U_2(t)\\
U_2(t+h) &=& A^{\otimes2} U_2(t)
\end{eqnarray}
Hence, the  Carleman matrix features two rows: $[A,B]$ and $[0,A^{\otimes2}]$.
Note that in general, $U_2$ contains non local terms of the form 
$u_i u_j$. Hence the number of Carleman 
variables at the second order truncation is $N_2 = O(G +G^2)$.

The efficiency of the procedure clearly depends on the level at which one
truncates the Carleman hierarchy, which in turn depends on the strength
of the nonlinearity and the desired time of evolution. In general, it is clear that each increasing step of the
ladder entails a corresponding increase of non-locality and a combinatorial
increase of dimensionality, meaning by this the number of Carleman variables.
This compounds with the fact that in a quantum computer it is highly unpractical to
measure the state $u(t+h)$, since this procedure has complexity $O(N_2^2)$.
One way out is to use telescopic time marching, namely
go from $u(0)$ to $u(t+T)$, $T=N_t h$, in a single "long-jump" update straddling
across $N)_t$ "small" steps.
This clearly aggravates the non-locality problem, since the
product $u_i(t+T)u_j(t+T)$ displays non-zero components across the entire 
grid (as soon as $T$ is comparable with the linear size $L$ of the domain).   

Details may change depending on the specific procedure, but these general
trends remain: transforming the nonlinearity away comes with 
a major computational burden.
Detailed experiments have shown that the Carleman-Navier-Stokes procedure 
exhibits very poor convergence even at moderate values of the Reynolds number. 
Typically, just a few steps with Carleman truncation below $K=4$
are found to incur errors well above 10 percent \cite{sanavio_three_2024}.

\section{Carleman Lattice Boltzmann}\label{sec:III}

A much better convergence has been found when the Carleman procedure is 
applied to the Lattice Boltzmann (LB) equation:
\begin{equation}
\label{eq:LBE}
f_p(\vec{x}+\vec{c}_p \Delta t ,t+\Delta t)-f_p(\vec{x},t) = -\omega [f_p(\vec{x},t)-f_p^{eq}(\vec{x},t)],\;p=1,\dots,b 
\end{equation}
where the discrete index $p$ labels the discrete velocities $\vec{c}_p$ and $\Delta t$ is the timestep.
The left side of the equation represents the {\it{streaming}} of the fluid towards the neighboring sites
along the $p$-th direction. 
Streaming is a nonlocal and linear operation which is exact on a classical computer as it does not involve
any floating-point operation since particles hop from one lattice site to the corresponding neighbor
with no loss of information.  
The right side of the equation represents the {\it{relaxation}} towards the 
local equilibrium, and $\omega$ is inversely proportional to the relaxation time-scale.
The parameter $\omega$ controls the LB fluid viscosity, hence the
Reynolds number,  according to
\begin{equation}\label{eq:Reynoldsomega}
\nu  = c_s^2 ({\frac{1}{\omega }-\frac{1}{2}}),
\end{equation}
$c_s$ being the lattice sound speed.

\begin{figure}[ht!]
\centering
\includegraphics[scale=0.5]{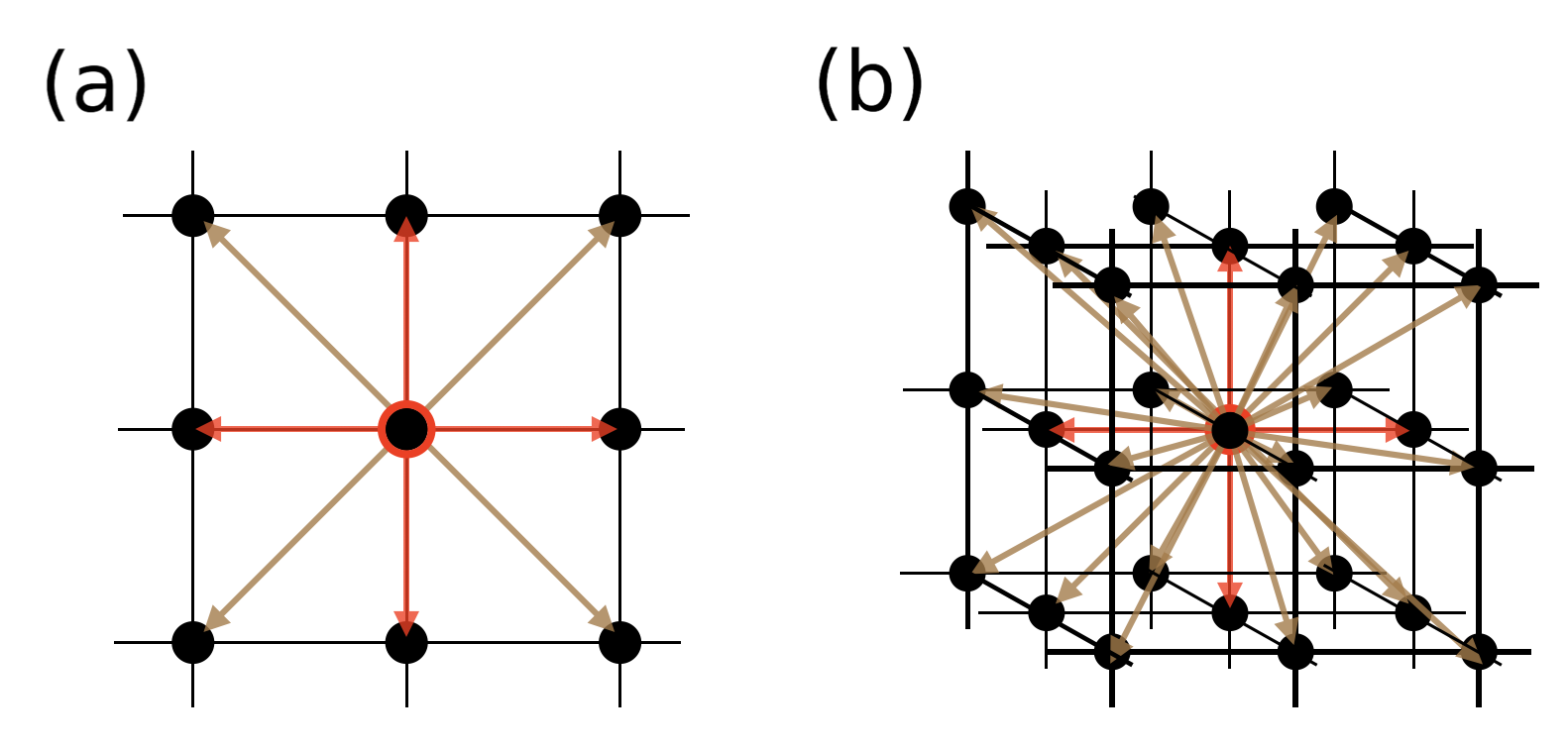}
\caption{(a) The D2Q9 model with $3^2=9$ velocities pointing to the neighboring sites, and  (b) the 
D3Q27 model with the $3^3=27$ velocities pointing to the neighboring sites in three-dimensions. 
They are generated by tensor product of the one-dimensional D1Q3 lattice with three discrete
velocities $(-1,0,+1)$.
The $0$-th velocity is represented by the red circle in the center of the lattice. \label{fig:LBModels}}
\end{figure}

The relaxation step is local and nonlinear, the nonlinearity of the fluid equations being 
entirely described by the quadratic terms in the local equilibrium: 
\begin{equation}\label{eq:equilibrium_LB}
f_p^{eq} = w_p (1 + \vec{u}_p + \frac{1}{2}(|\vec{u}_p|^2 -|\vec{u}|^2))
\end{equation}
where $\vec{u}_p \equiv \frac{\vec{u} \cdot \vec{c_p}}{c_s^2}$, and $w_p$ being  
a weight corresponding to the equilibrium distribution with no flow ($\vec{u}=0$). 
The values of the weights $w_p$ depend on the specific model, popular versions being 
the D2Q9 (two dimensions, 9 velocities) and the D3Q27 (three dimensions, 27 velocities). 
Those are represented in Fig.~\ref{fig:LBModels}.

% -------------------------------------------

The Carleman Lattice Boltzmann (CLB) procedure consists in
treating the second (and higher) order terms $f_{pq}= f_pf_q$, coming from 
the expansion of Eq.~\eqref{eq:equilibrium_LB}, as new variables and then 
defining their evolution following Eq.~\eqref{eq:LBE}. 
This can be written as a linear system in the form
$$
F(t+\Delta t)=\mathcal{C}F(t),
$$
\noindent where $F(t)$ is the vector collecting the first $f_p$ and second $f_{pq}$ order terms 
at all lattice sites while $\mathcal{C}$ is the Carleman matrix.

In the LB method, we can treat streaming and relaxation separately and the same holds
true for the CLB method, namely:
$$
\mathcal{C} = \mathcal{S} \mathcal{R}.
$$
The streaming matrix $\mathcal{S}$ is unitary, while the relaxation matrix $\mathcal{R}$
is not, as it encodes the irreversible decay to the local equilibrium
leading to emergent dissipation. 
Thus we can separate the two processes and apply streaming after relaxation. 

The streaming operator for the D2Q9 model has been thoroughly investigated 
in~\cite{sanavio_lattice_2024}, but being unitary, it does not entail the same burden 
as the implementation of the nonunitary operator $\mathcal{R}$. 

In fact, it can be expressed by a circuit whose depth scales quadratically 
with the number of qubits, and a width that is twice the original~\cite{barenco_1995}.

The CLB relaxation process, truncated at second order, turns into the following linear system

\begin{eqnarray}
f_p(x+hc_p,t+h) &=& A_{pq}f_q(x,t)+B_{pqr}f_{qr}(x,t)\\
f_{pq}(x+hc_p,y+hc_q,t+h)&=&A_{pr}A_{qs}f_{rs}(x,y,t),
\end{eqnarray}

\noindent This allows to write the relaxation matrix over $N$ lattice sites as

\begin{equation}\label{eq:CLB_matrix}
\mathcal{R}_N = \begin{pmatrix}
1_N\otimes A & \Delta\otimes B\\
0 & 1_{N^2}\otimes A^{\otimes 2}
\end{pmatrix},
\end{equation}

\noindent where the matrix $\Delta$ accounts for the locality of the relaxation process. 
Thus the non-null components $\Delta_{x,y}=\delta_{x,(N+1)y}$ with $x,y= 0,\dots,N-1$,
couple together the local Carleman variables of first and second order.
For notational simplicity, in the following the spatial indices will be denoted simply 
by $i,j,k$. 
 
\begin{figure}[ht!]
\centering
\includegraphics[scale=0.5]{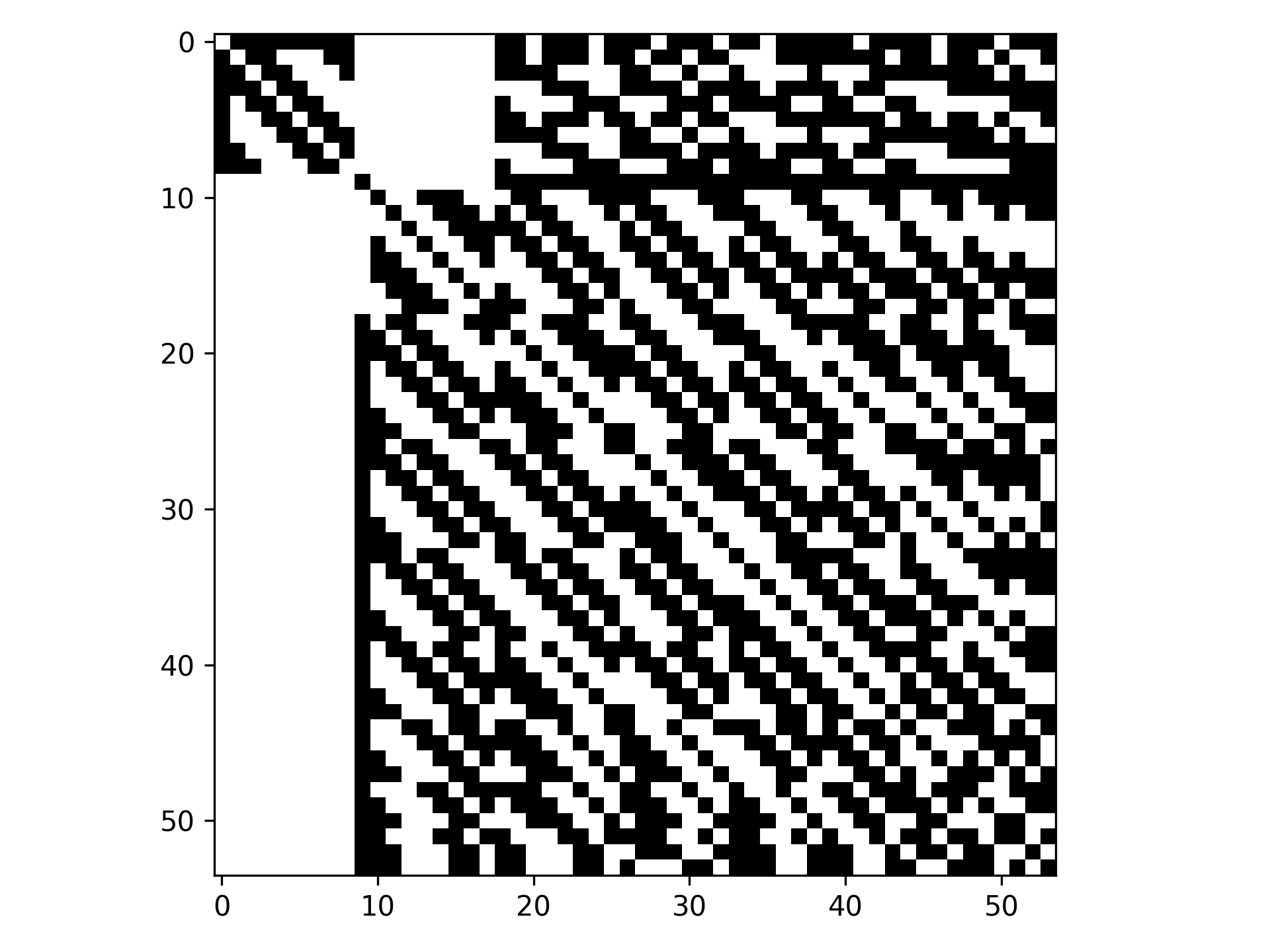}
\caption{A black-and-white representation of the Carleman Lattice Boltzmann 
relaxation matrix $\mathcal{R}_1$ for a single lattice site for the D2Q9 model. 
In the white blocks, the value of $\mathcal{R}_1$ is 0. \label{fig:CLB_matrix}}
\end{figure}

The matrix $B$ gives the strength of the coupling between first and second order 
Carleman variables. Note that this is analogous to Eq.~\eqref{C2}, but we have 
detailed the dependence on the position of the variables. 
The single lattice-site relaxation matrix for the D2Q9 model takes the form 
plotted in Fig.~\ref{fig:CLB_matrix}. 
The block in the upper-left side of the matrix is the b/w representation 
of the matrix $A$, which is a $9\times9$ full matrix. 
The large $81\times81$ matrix in the bottom-right side of the matrix represents the matrix $B$. 
Although this matrix has a sparsity $s=81$, out of 90 rows and columns, it is worth to notice that 
its multiple lattice-sites version $\mathcal{R}_N$ (Eq.~\eqref{eq:CLB_matrix}) has the same sparsity, regardless of the number of lattice sites $N$. For any LB model with $b$ velocities, the sparsity of the Carleman relaxation matrix truncated at order $\tau$ is $s=b^\tau$.

% -------------------------------
% --------------------------------------------------------------

\begin{figure}[ht!]
\centering
\includegraphics[scale=0.49]{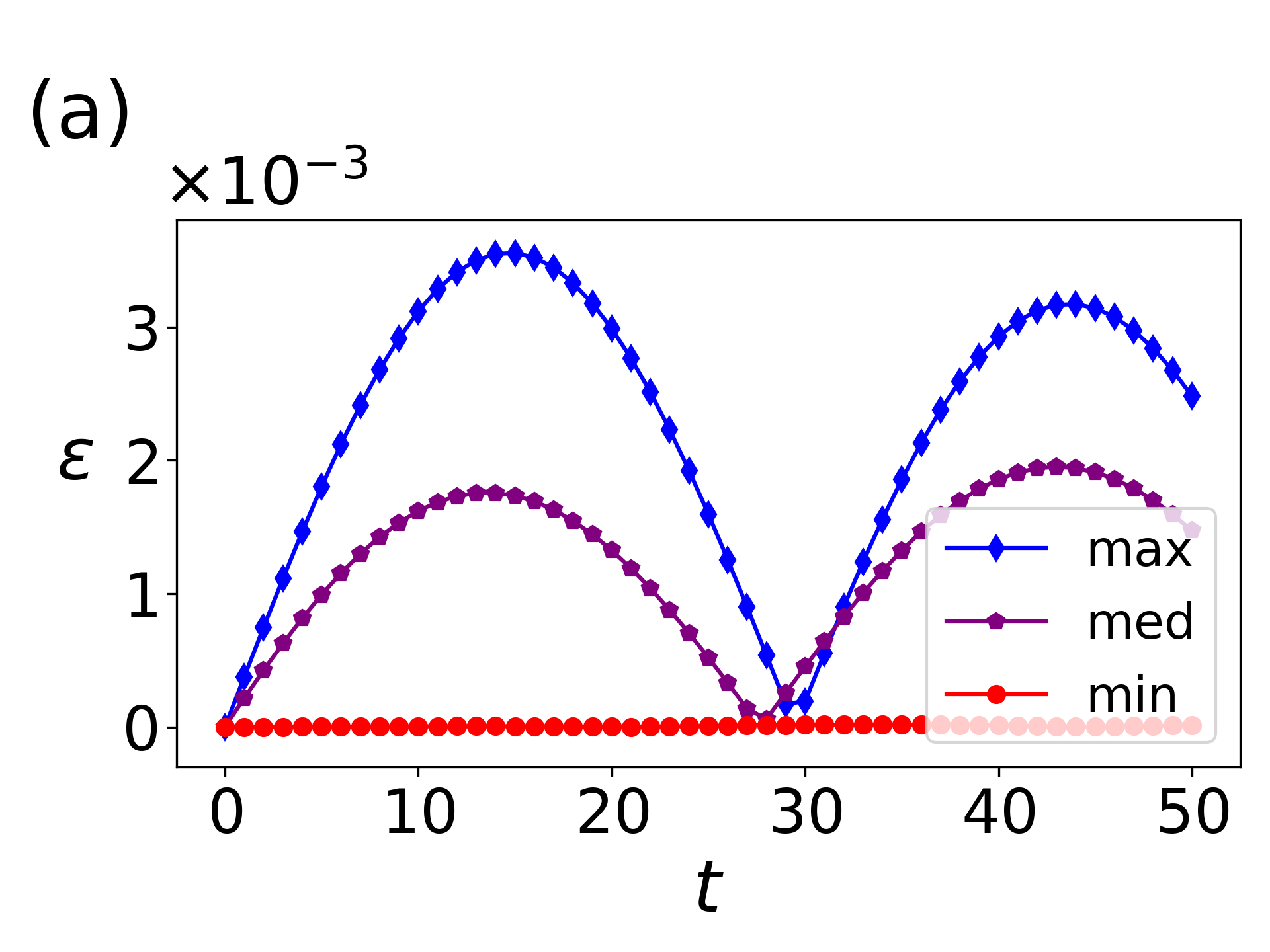}
\includegraphics[scale=0.49]{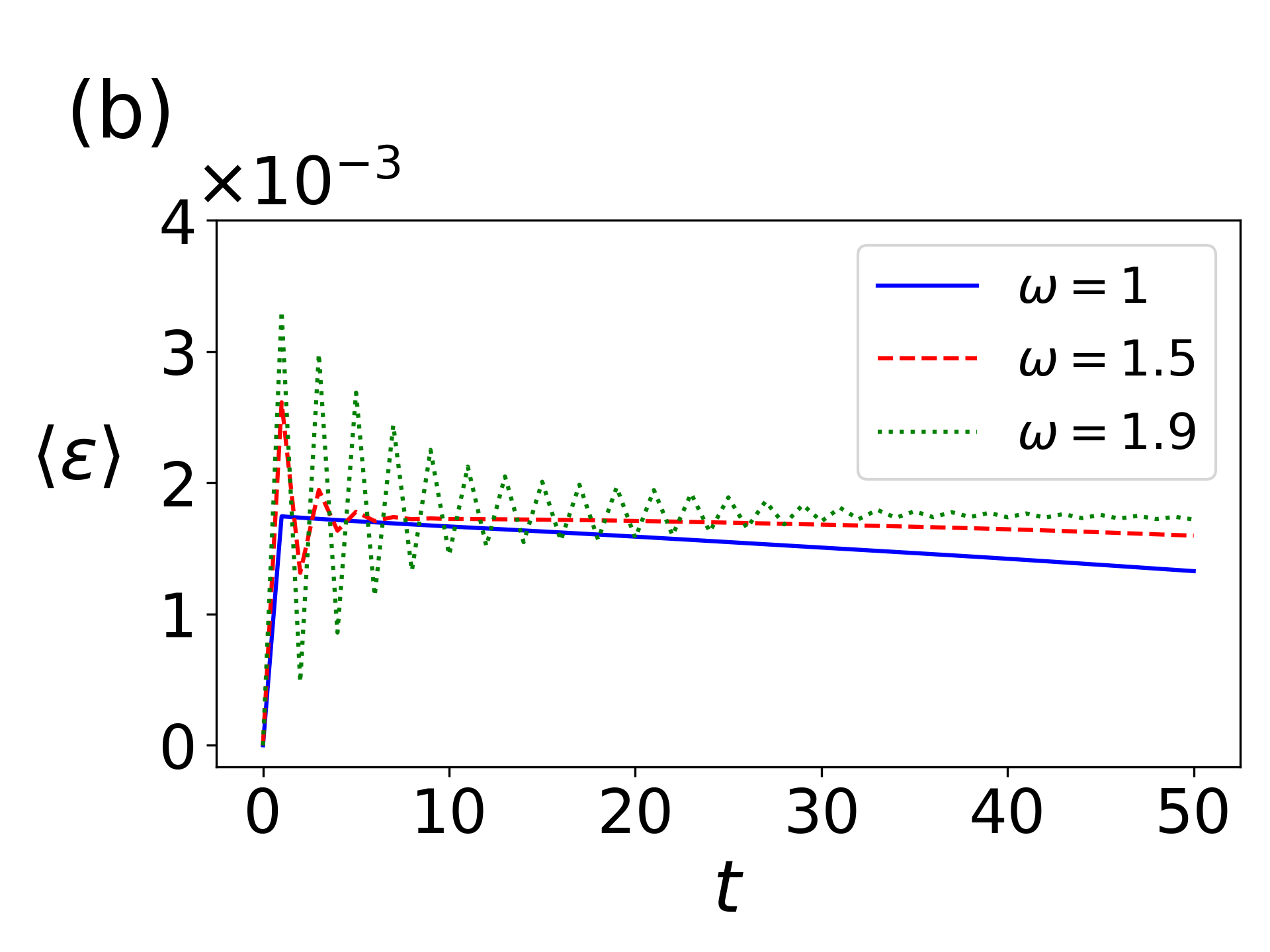}
\caption{In~(a) the relative error $\epsilon$ between the results obtained by LBM and by Carleman linearization 
truncated at second order for a Kolmogorov-like flow simulated on a $48\times48$ square lattice.
The error is monitored at three points, where $\epsilon$ is maximum (blue diamonds), median 
(purple pentagons) and minimum (red circles). 
In (b) the error averaged over all the lattice sites  for different values of $\omega$. \label{fig:Carleman_Kolmogorov}}
\end{figure}

Since dissipation is emergent and does not require any Laplacian, the 
nonlinear/linear ratio in this formulation amounts to comparing linear 
and quadratic terms in the local equilibria, which is simply given by the Mach number.
This makes a huge difference as compared to Navier-Stokes since
the Mach number is generally orders of magnitude smaller 
than the Reynolds number.
Hence, LB presents a much smaller nonlinearity barrier to the 
Carleman procedure \cite{li_potential_2023}.
This intuition is indeed backed up by numerical experiments, which show
excellent convergence of the CLB procedure. Even at second order truncation,
flows with Reynolds number $O(100)$ exhibit relative errors of 
about $10^{-3}$ over hundreds of timesteps in the case of a 2D Kolmogorov-like flow~\cite{sanavio_lattice_2024} calculated on a $48\times 48$ regular lattice, and simulated with lattice units $\Delta x = \Delta t = 1$ and $c_s=1/\sqrt{3}$.
In Fig.~\ref{fig:Carleman_Kolmogorov}(a) we show the relative error $\epsilon$  of the CLB method 
for the three "max-med-min" points of the lattice where it reaches the maximum, median and minimum value. 

In Fig.~\ref{fig:Carleman_Kolmogorov}(b) we show the relative error averaged over the whole lattice $\langle \epsilon\rangle$. The curves refer to different values of the Reynolds number, $\text{Re}\approx 30,90,550$ obtained by varying the parameter $\omega = 1,1.5,1.9$ respectively, with $|\vec{u}|\approx0.1$, see Eqs.~\eqref{eq:Reynoldsomega} and \eqref{eq:Reynolds}.

Much more systematic work is needed to assess whether the very encouraging convergence 
properties of the CLB procedure extend to more complex flows, such as high-Reynolds numbers
in confined geometries. To this regard, it is important to observe that such an analysis can only be
performed on a quantum computer, since the CLB procedure rapidly saturates the memory capacity
of classical ones, due to the exponential growth of the number of Carleman variables with the grid size and
the truncation level of the Carleman embedding.  

In the following, we address precisely this issue.

\section{Towards a quantum CLB algorithm}\label{sec:IV}

\begin{figure}[ht!]
\centering
\includegraphics[scale=0.49]{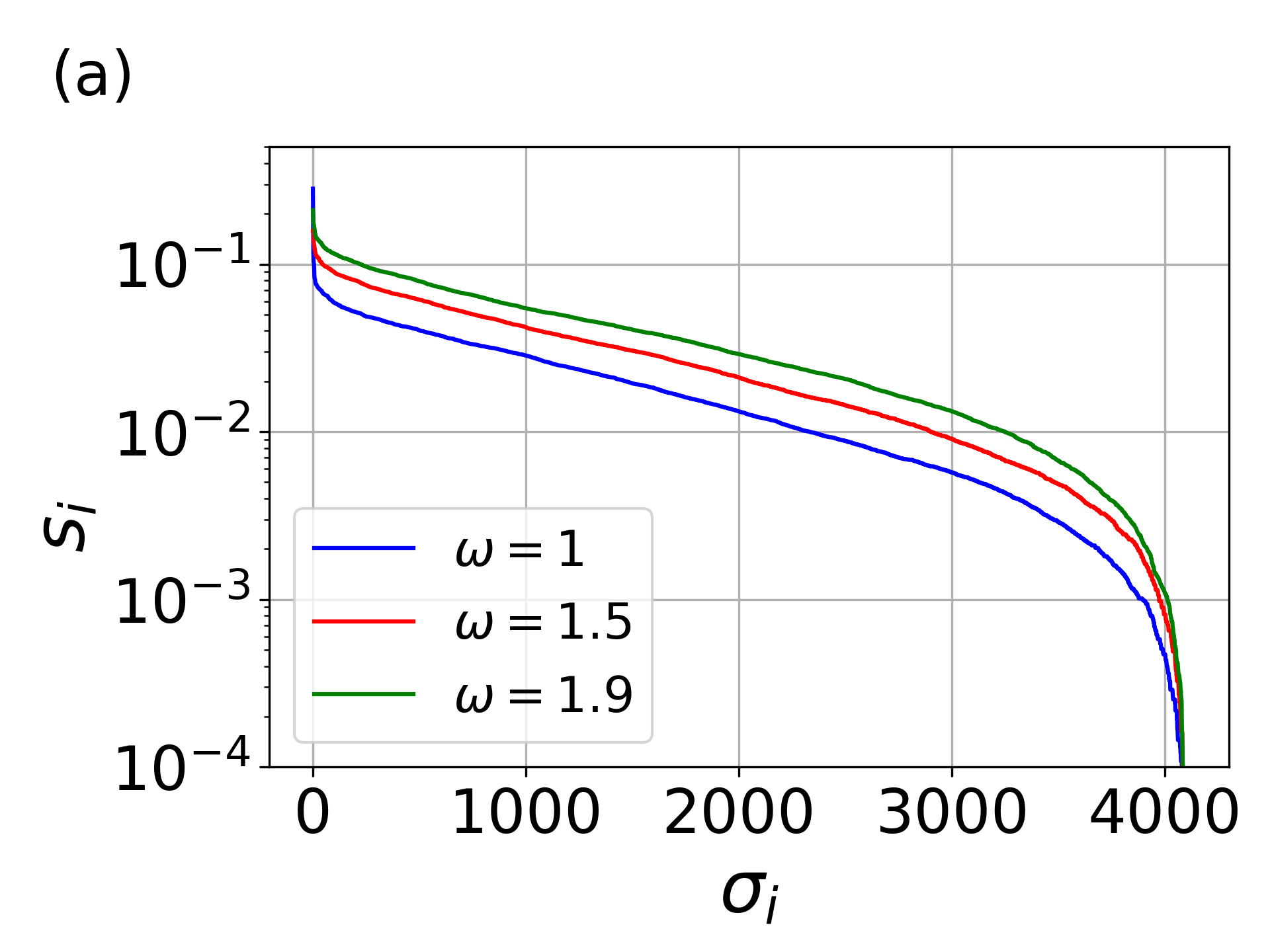}
\includegraphics[scale=0.49]{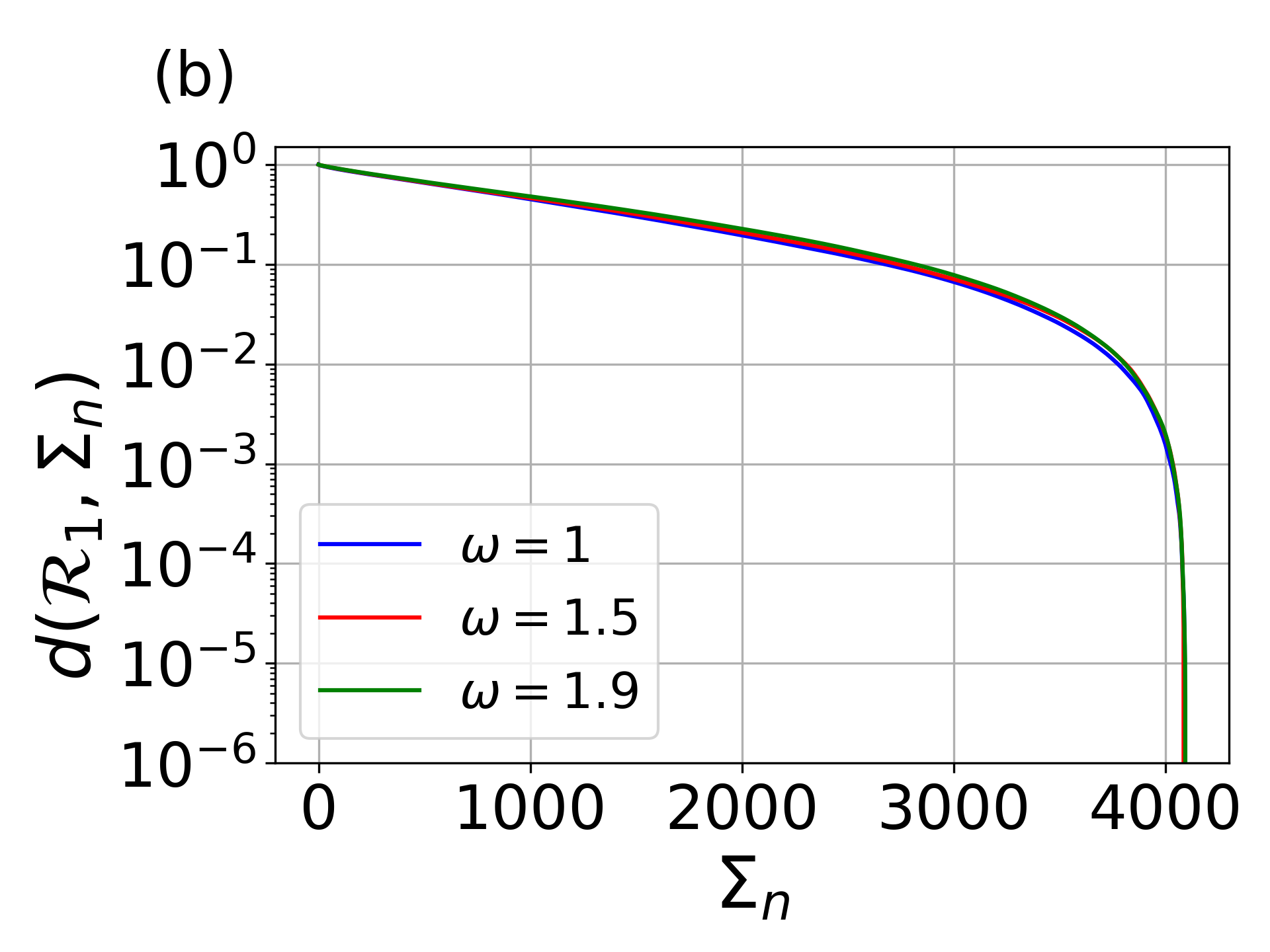}
\caption{The expansion of the single lattice-site Carleman matrix $\mathcal{R}_1$. In~(a) the coefficients $s_i$ of the expansion for each basis matrix $\sigma_i$. In~(b) the normalized Frobenius distance $d$ between matrix $\mathcal{R}_1$ and the approximation $\Sigma_n$ obtained with the first $n$ elements of the expansion.\label{fig:Pauli_expansion}}
\end{figure}

The development of a  quantum CLB algorithm faces a number of challenges. 
In a nutshell, the main problem is that even the single-lattice site
CLB matrix projects over the entire series $\{\sigma_i\}_{i}$ of tensor Pauli 
gates, obtained by all the possible combinations of tensor product between the three 
Pauli matrices and the identity. 
In Fig.~\ref{fig:Pauli_expansion}(a) we plot the value $s_i$ of the associated coefficients sorted 
in decreasing order. To evaluate the performance of the expansion we calculate the normalized Frobenius distance 
\begin{equation}
d(\mathcal{R}_1,\Sigma_n) = \frac{||\mathcal{R}_1-\Sigma_n||}{||\mathcal{R}_1||}
\end{equation} 

\noindent between the Carleman matrix $\mathcal{R}_1$ detailed in Eq.~\eqref{eq:CLB_matrix} and the matrix obtained by partial expansion of $\mathcal{R}_1$ with the most relevant $n$ elements $\Sigma_n = \sum_{i=1}^ns_i\sigma_i$. We plot it in Fig.~\ref{fig:Pauli_expansion}(b).

% --------------------------------
% -------------------------------

As a result, standard techniques usually employed for Hamiltonian simulation such as low-order 
Trotterization~\cite{whitfield_simulation_2011} or linear combination of 
unitaries~\cite{childs_hamiltonian_2012,mezzacapo_quantum_2015} 
imply a depth of the associated quantum circuit that scales exponentially
with the number of qubits~\cite{sanavio_lattice_2024}, making the CLB scheme practically unviable. 
In the following we present a circuit for the CLB that overcomes this problem
by employing block-encoding techniques \cite{low_optimal_2017} for sparse operators.

\section{Block-encoding oracles}\label{sec:V}

The depth problem of the circuit can be circumvented by using block-encoding for block-sparse matrices, namely by exploiting the sparse-matrix representation ($i,pos(i),v_{i,pos(i)}$), which encodes for each row $i$ the position $pos(i)$ of the non null values $v_{i,pos(i)}$ for the matrix $\mathcal{C}$. 
With the block-encoding, we can define the non-unitary matrix $\mathcal{C}$ as a block of a larger unitary operation $U(\mathcal{C})$, such that

\begin{eqnarray}\label{eq:block_encoding}
U(\mathcal{C}) = \begin{pmatrix}
\mathcal{C}/\gamma & \star\\
\star & \star
\end{pmatrix},
\end{eqnarray}

\noindent where $\gamma = \max_{i,j}[v_{i,j}]$ is the maximum value in $\mathcal{C}$, while the symbol $\star$ represents elements that we can ignore, provided they satisfy the condition for $U$ to be unitary.
Thus, when applied to a state $|\psi\rangle$, the (normalized) evolution $\mathcal{C}|\psi\rangle$ is obtained when the ancilla qubits used for the block-encoding are measured in the state $|0\rangle$.

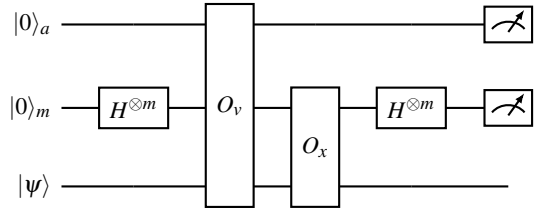
\begin{figure}[t]
  \centering
\begin{quantikz}
\lstick{$|0\rangle_a$}  &  \qw &  \gate[3]{O_v}  &  \qw&\qw& \meter{} \\
\lstick{$|0\rangle_m$}  & \gate{H^{\otimes m}}&               & \gate[2]{O_x}&  \gate{H^{\otimes m}} & \meter{} \\
\lstick{$|\psi\rangle$} & \qw                 &               &	             &                   \qw & \qw
\end{quantikz}  
\caption{The circuit for the implementation of the dynamics using the block-encoding technique and matrix access oracles. 
Given a matrix $\mathcal{C}$, the operator $\hat{O}_x$ is the position oracle which embeds the position of the 
non zero elements while $\hat{O}_v$ is the value oracle which registers their values. 
If all the ancilla qubits are measured in the state $|0\rangle$, the state is proportional to $\mathcal{C}|\psi\rangle$, while
any other combination leads to a different state, hence to an error.}
  \label{fig:oracle_block_encoding}
\end{figure}
% ---------------------------  EFIG

The block-encoding strategy~\cite{low_optimal_2017}\cite{camps_explicit_2024} requires the use of two extra quantum registers, labeled $m$ and $a$, initialized in the state $|0\rangle$ and the use of two oracles operators, $\hat{O}_x,\hat{O}_v$ as described next. 
The quantum register $|0\rangle_m$ consists of $m$ ancilla qubits, where $m =\log_2 s$, $s$ being 
the sparsity of the matrix (the maximum number of non-zero elements per row, for all rows in the matrix). 
It is used to store the position $j=pos(i)$, for each non zero entry $\mathcal{C}_{i,j}$ along the $i-th$ row. 
Thus the oracle $\hat{O}_x$ on the state $|i\rangle$ acts as follows:

\begin{equation}
\hat{O}_x|i\rangle|0\rangle_m= |i\rangle|pos(i)\rangle_m.
\end{equation}

A second oracle $\hat{O}_v$, provides the value $\mathcal{C}_{i,pos(i)}$, by applying a controlled rotation 
onto an extra ancilla qubit:

\begin{equation}
\hat{O}_v|i\rangle|j\rangle_m|0\rangle_a= |i\rangle|j\rangle_m(\mathcal{C}_{i,j}|0\rangle_a+\sqrt{1-\mathcal{C}_{i,j}^2}|1\rangle_a).
\end{equation}

\noindent The circuit is plotted in Fig.~\ref{fig:oracle_block_encoding}.

The oracles take full advantage of the sparse nature of 
the Carleman matrix, which is particularly useful whenever the sparse matrix presents 
characteristics such as sorted position and values repetitions. 
For instance, a tridiagonal Toeplitz matrix 
can be mapped into a quantum circuit with just a
polynomial depth.~\cite{camps_explicit_2024,sanavio_explicit_2024}. 

This advantage comes at a cost, though. The nonunitary matrix $\mathcal{C}$ is implemented only on 
the subspace where all the ancilla qubits are in the state $|0\rangle$, thus making 
the circuit probabilistic, with a success rate $p_s$ that depends on the degree of non-unitarity 
of the Carleman matrix, as well as on the initial state $\sum_i c_i|i\rangle$. 
In fact, the equation of the success rate takes the form~\cite{sanavio_explicit_2024}

\begin{equation}
p_s = \frac{1}{2^{2m}}\lVert\sum_{i,l}c_i\mathcal{C}_{i,c(i,l)}|c_i,l\rangle\rVert^2.
\end{equation}

However, the dominant contribution is the one due to the number of ancilla qubits used for 
the embedding, which dictates a success rate $p_s\approx\frac{1}{2^{2m}}$. 
In our case, $m$ is proportional to the sparsity, which is given by the square of 
the number of discrete velocities $s=b^2$, leading to:
$$
p_s\approx\frac{1}{2^{4\lceil{\log_2b}\rceil}} \approx\frac{1}{b^4}.
$$

\noindent Taking into account the ceiling factor, this yields 
$p_s\approx 4 \cdot10^{-3}, 6 \cdot 10^{-5} , 4 \cdot 10^{-6}$ for the 
D1Q3, D2Q9 and D3Q27 LB models, respectively. 

For the D2Q9 model defined on a lattice with $N$ sites, the relaxation matrix expressed in Eq.~\eqref{eq:CLB_matrix} has a sparsity $s=81$, due to the second order term $A\otimes A$. Note indeed that both $A$ and $B$ are quite full, since  $A$ itself is a $9\times 9$ matrix and $B$ is a $9\times 81$ matrix. However, the global sparsity of $\mathcal{R}$, which is a 
square operator with $9 N+81 N^2$ rows, remains $81$ regardless of the total number of  lattice sites $N$. 

We can exploit this property by using the matrix access oracles, that enable us to encode 
$\mathcal{R}_N$ with $m=7$ ancilla qubits, which results in a total number of gates 
scaling linearly with the size of the lattice.

\begin{figure}[t]
  \centering
  \includegraphics[scale=0.35]{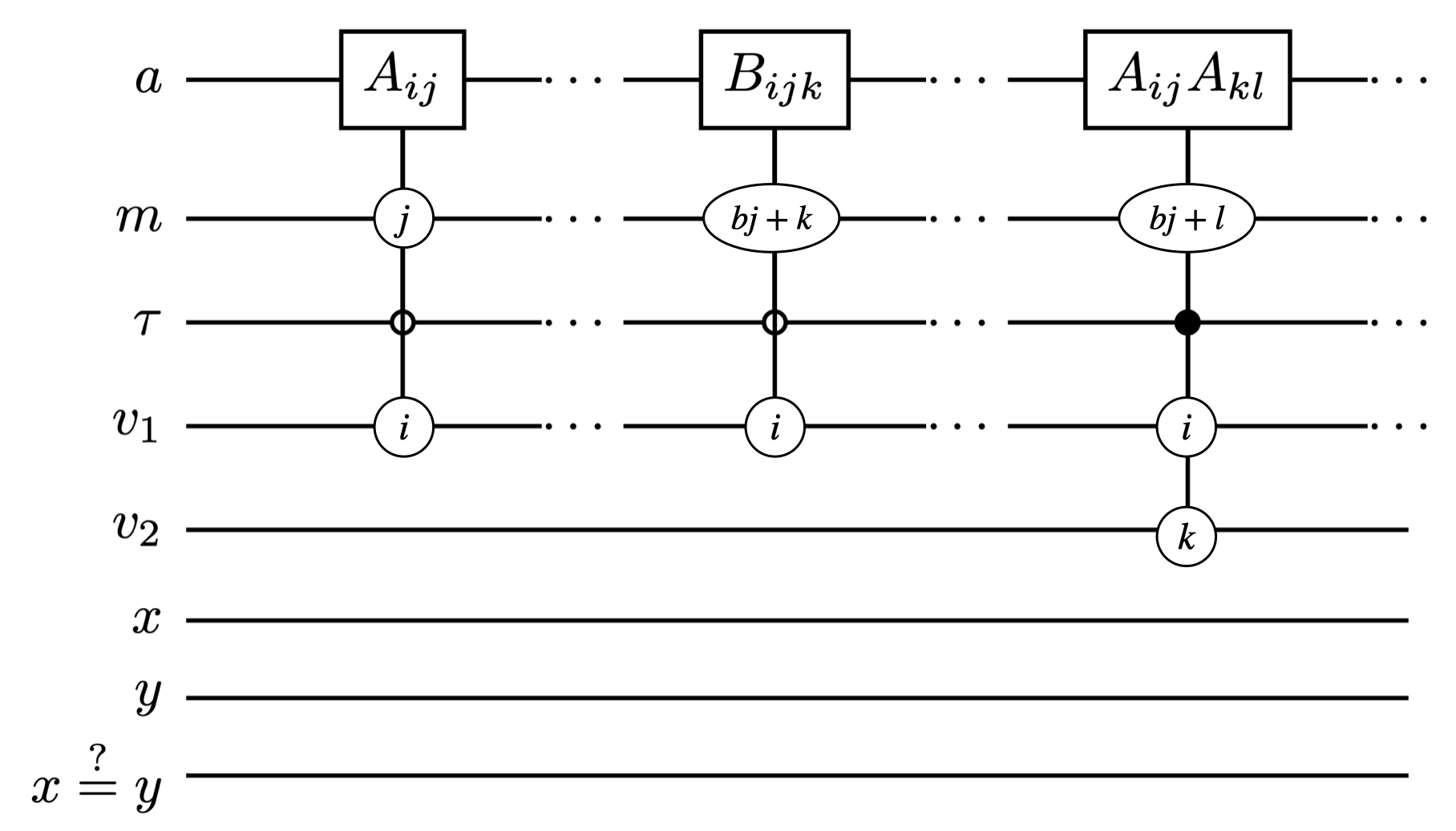}
\caption{The value oracle $\hat O_v$ for the D2Q9 LBM truncated at second order. 
A controlled rotation applies the value $A_{ij}$ on the ancilla qubit $a$ if the state in $m$ is $|i\rangle$ the state in $\tau$ is $|0\rangle$ and the state in $v_1$ is $|j\rangle$. The same happens for the $B_{ijk}$ and the $A_{ij}A_{kl}$ matrices. In this figure we followed the convention introduced in~\cite{sanavio_lattice_2024}, where a circle including a number $n$ implies a control conditioned by the state $|n\rangle.$}
  \label{fig:LBM_oracle_circuit_a}
\end{figure}

\begin{figure}[t]
  \centering
    \includegraphics[scale=0.35]{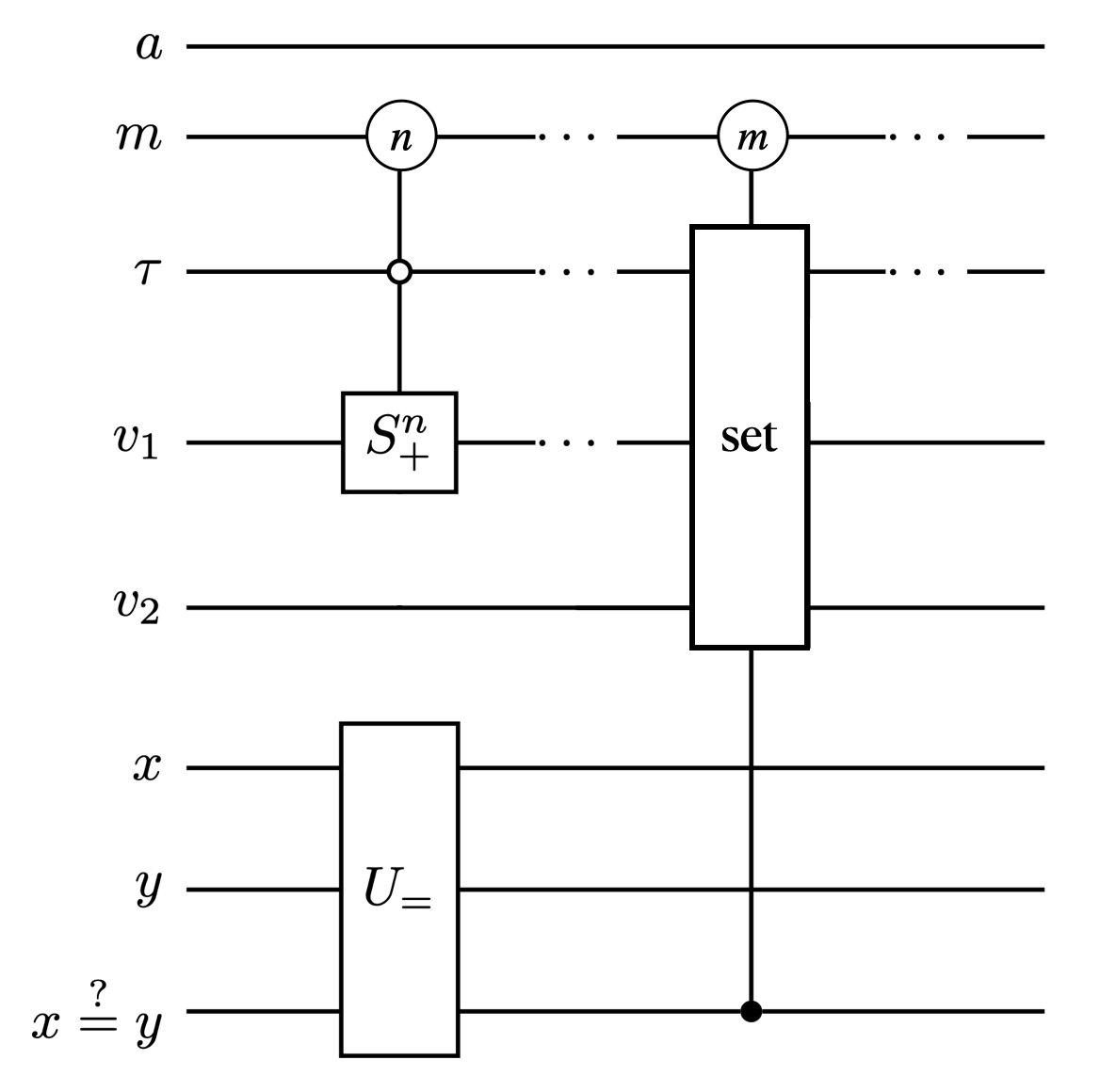}
%\begin{quantikz}
%\lstick{$a$}   & \qw & \qw        &	 \qw       & \qw          & \qw       &\qw \\
%\lstick{$m$}   & \qw & \octrl{1} & \qw \dots\ & \octrl{1}   &\qw \dots\     &\qw \\
%\lstick{$\tau$}& \qw &  \octrl{2}& \qw \dots\ & \gate{X_0}   &\qw \dots\   &\qw \\
%\lstick{$v_1$} & \qw &  \gate{S_+^n}& \qw \dots\ & \gate[2]{set}   &\qw  &\qw \\
%\lstick{$v_2$} & \qw & \qw        &	 \qw       & \qw          &\qw         &\qw \\
%\lstick{$x$}   & \qw &\gate[3]{U_{=}}&	 \qw   & \qw          & \qw       &\qw \\
%\lstick{$y$}   & \qw &            &	 \qw       & \qw          & \qw       &\qw \\
%\lstick{$x\overset{?}{=}y$}&\qw&  & \qw      & \ctrl{-5}      & \qw   &\qw \\
%\end{quantikz}  
\caption{The position oracle $\hat O_x$ for the D2Q9 LBM truncated at second order. Conditioned on the state of the $m$ register $|n\rangle$, for $n=0,1,\dots,s$, we apply the shift operator $S_+^n$, which finds the position of the $n-th$ element of the relaxation matrix $\mathcal{R}$. We do it both for $\tau=|0\rangle$ and $|1\rangle.$ For $m=0,1,\dots,s $ and conditioned on the state $x=y$ being $|1\rangle$, we apply the $set$ operator on the registers $\tau,v_1,v_2$, as described in the main text.}
  \label{fig:LBM_oracle_circuit_b}
\end{figure}

The circuit acts on the quantum registers shown in Figs.~\ref{fig:LBM_oracle_circuit_a} and \ref{fig:LBM_oracle_circuit_b}, where $a$ and $m$ 
are the ancilla qubits used for the block-encoding, $\tau$ is a single qubit accounting for 
the truncation order, $v_1,v_2$ are the velocity registers with $\lceil \log_2b\rceil$ qubits and $x,y$ the position registers with $q_N=\lceil \log_2N\rceil$ qubits. By using these quantum registers, the first order functions are encoded in the state $\sum_{i,x}f_i(x)|0\rangle_\tau|i\rangle_{v_1}|x\rangle_x$, whereas the second order functions are encoded in the state $\sum_{i,j}\sum_{x,y}f_i(x)f_j(y)|1\rangle_\tau|i\rangle_{v_1}|j\rangle_{v_2}|x\rangle_x|y\rangle_y$, both properly normalized and while the remaining qubits are in the state $|0\rangle$ if not specified otherwise.

A last single ancilla qubit is switched on when $x=y$, which is implemented through a 
comparator circuit $U_{=}$, which sets the $x=y$ ancilla qubit to $|1\rangle$ if $|x\rangle=|y\rangle$. 
The linear dependence of the depth with $\log_2N$ for the relaxation operator comes from this comparator circuit. 
Fig.~\ref{fig:LBM_oracle_circuit_a} shows the circuit for the values oracle. 
This applies controlled rotations on the value qubit $a$ depending on the sparsity 
index $m$, the truncation order $\tau$ and the velocities $v$. 

\noindent Fig.~\ref{fig:LBM_oracle_circuit_b} shows the circuit for the position oracle. 
This is more cumbersome than the values oracle, as it uses different operators. 
$S_+$ is the rightward shift operator, and it is used to find the position 
of the diagonal blocks of $\mathcal{R}_N$, namely $A$  and $A^{\otimes 2}$. 
The $\text{set}$ operator accounts for the $B$ on the off-diagonal blocks of $\mathcal{R}_N$,
which comprises a limited number of CNOTs and single qubits rotations. 
This number is fixed, regardless of the number of gridpoints.

Hence, the circuit for implementing the relaxation term $\mathcal{R}_N$ has a linear dependence with $q_N$, which is given solely by the comparator circuit $U_{=}$. At the same time, the circuit for implementing the streaming operator $\mathcal{S}$ has a depth that is quadratic with $q_N$~\cite{barenco_elementary_1995,sanavio_lattice_2024}, thus making the whole circuit gate complexity $\mathcal{O}(q_N^2)$.

% --------------------------------------
% ---------------------------  EFIG

% ---------------------------  EFIG

\noindent We remind that the success probability of implementing a single time step for the D2Q9 model lies in 
the order of $p_s\approx 10^{-5}$. 
We performed the calculation to retrieve the exact probability distribution 
in the case of an initial state where all the velocity
distributions are uniform, $f_i = \frac{1}{9}$, and in the case of no-flow equilibrium, with $f_i=w_i$ for $i=0,\dots,8.$
%
%\begin{eqnarray}
%|\psi_{\text{eq}}\rangle &=&|0\rangle_\tau\sum_{i=0}^{b-1}\sum_{x=0}^{N-1}\frac{1}{\sqrt{2Nb}}|i\rangle_{v_1}|0\rangle_{v_2}|x\rangle_x|0\rangle_y\\
%&&+,|1\rangle_\tau\sum_{i,j=0}^{b-1}\sum_{x,y=0}^{N-1}\frac{1}{\sqrt{2N^2b^2}}|i\rangle_{v_1}|j\rangle_{v_2}|x\rangle_x|y\rangle_y.
%\end{eqnarray}

\begin{figure}[ht!]
\centering
\includegraphics[scale=0.35]{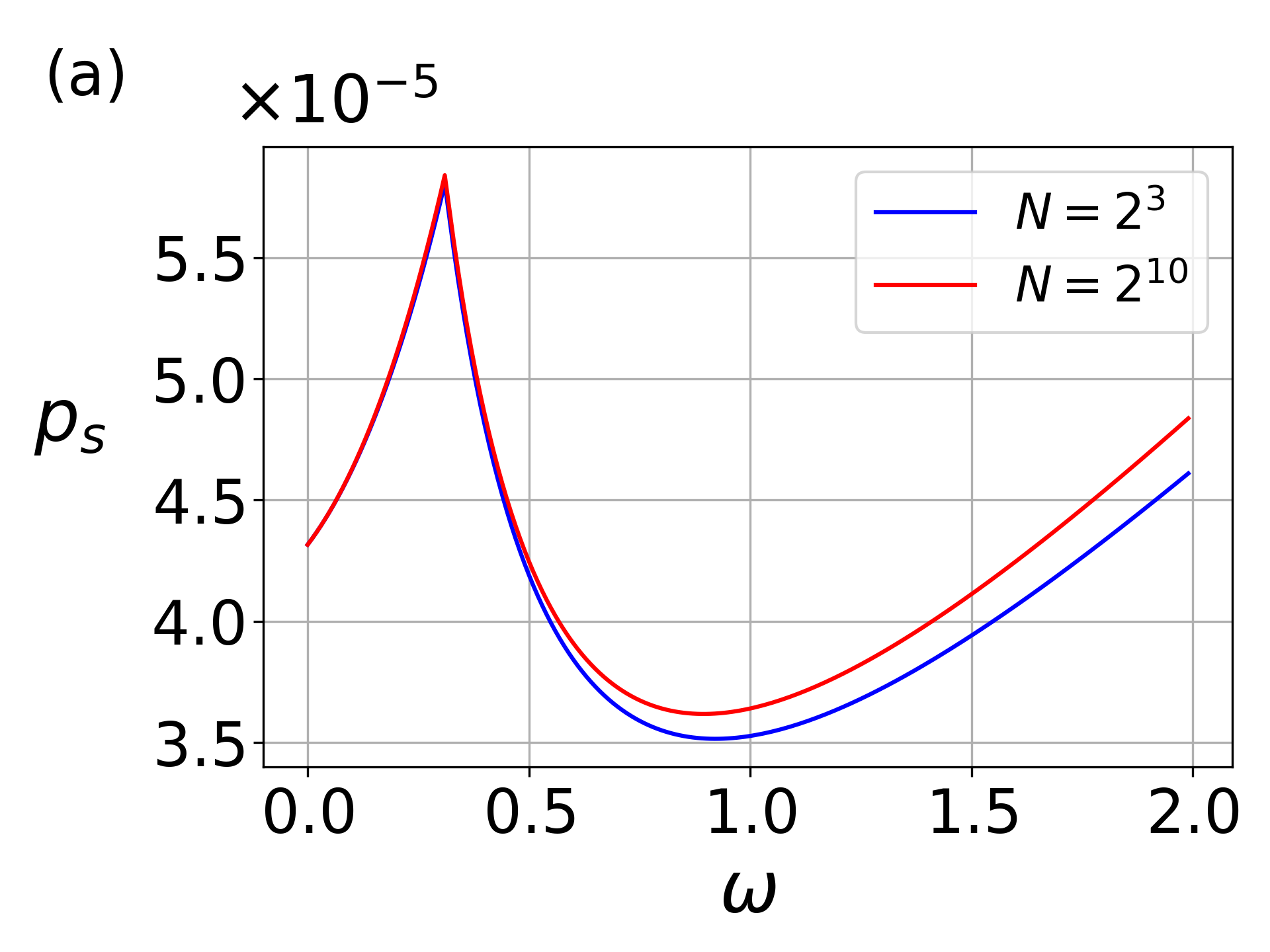}
\includegraphics[scale=0.35]{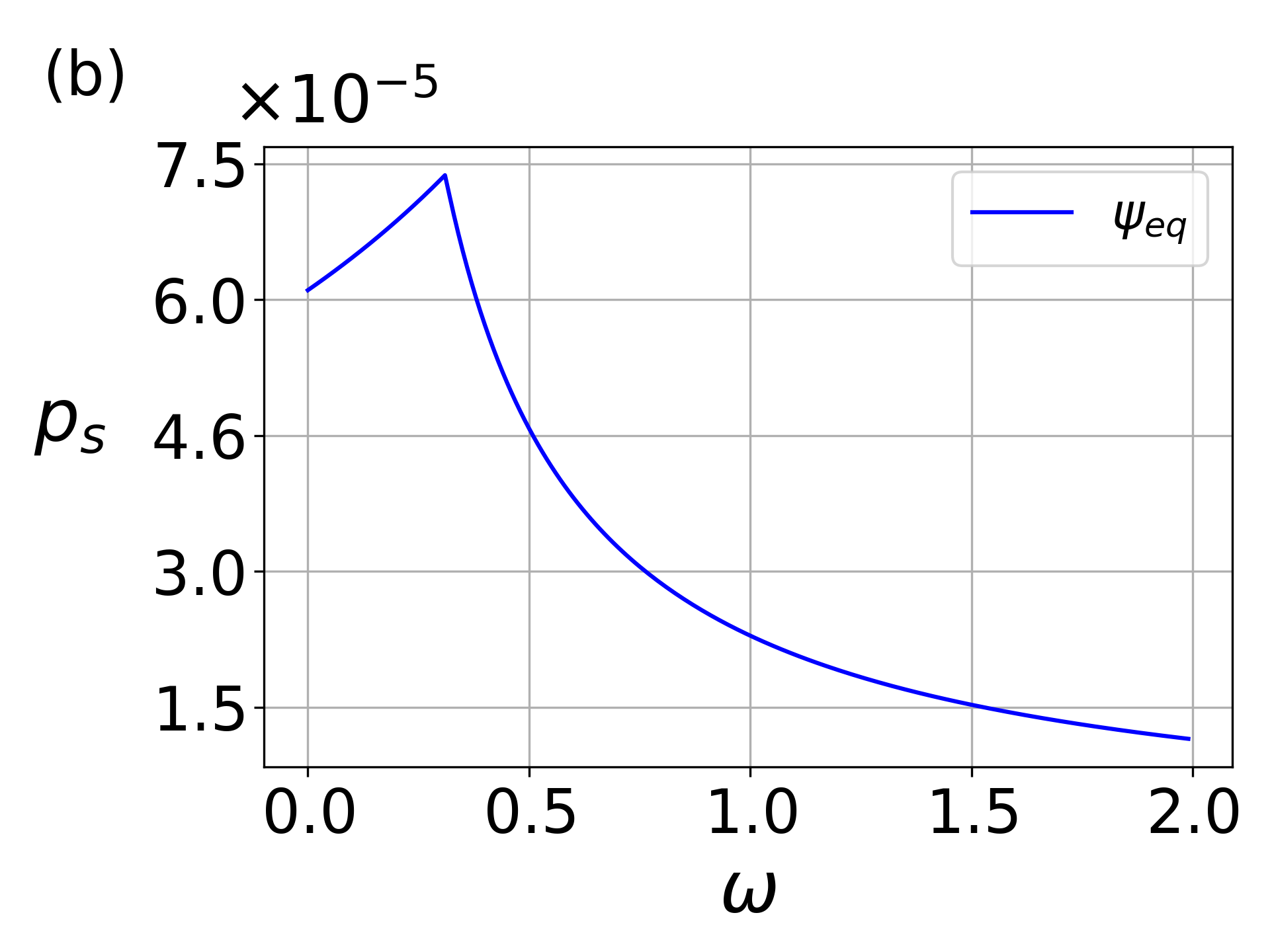}
\caption{The success probability $p_s$ of a single time step by using the block encoding technique as a function of $\omega$ obtained for (a) the uniform intial conditions $f_i=1/9$ for each site, and (b) the equilibrium condition $f_i=w_i$ for each site. \label{fig:success_probability}}
\end{figure}

\noindent We found that in both cases the probability distribution has a maximum at $\omega = 0.31$, as shown in Fig.~\ref{fig:success_probability}. This curve exhibits a cusp that stems from the renormalization that we need to apply to 
the matrices $A$ and $B$, since the block-encoding technique is restricted to matrices with maximum values equal to 1, cf. Eq.~\eqref{eq:block_encoding}. 
Fig.~\ref{fig:success_probability}(a) shows the success probability for a single time-step for the uniform case and a number of lattice sites $N=2^3$ (blue curve) and $N=2^{10}$ (red curve). For higher values of $N$, the success probability is similar to the one obtained for the case $N=2^{10}$.
Fig.~\ref{fig:success_probability}(b) shows the success probability for the equilibrium case. 
Here the result does not depend on the number of gridpoints, as it is expected from the equilibrium condition. 

% ---------------------------------------------------------------------

The use of block-encoding with oracles can provide an efficient circuit in terms of number of 
two-qubit gates necessary to implement a single time step.
Clearly those results are shaded by the low value of the success probability. 
In the lookout and conclusion section we discuss some ideas that may help mitigating this problem
by increasing the amplitude of the desired state.

\section{Conclusions and future outlook}\label{sec:VI}

Summarizing, the classical implementation of the CLB procedure shows 
excellent convergence, at least for the 2D Kolmogorov flow discussed in this paper.
In the quantum realm, a naive projection of 
the CLB matrix into the Pauli quantum gates faces with an exponential depth problem.
Such an exponential depth can be brought down to a quadratic one by using
block encoding oracles for block-sparse matrices. 
In this work we have developed the explicit form of the corresponding quantum circuit.
However, this technique requires ancilla qubits, thus turning the quantum algorithm 
into a probabilistic one, whose success probability decays as the inverse of the forth power of the number of discrete velocities. 
For the case of the D2Q9 lattice this leads to a single-step success probability
$p_s \sim 10^{-5}$, which makes the algorithm unviable on present day quantum computers. 
On the other hand, this probability is nearly independent of the number of lattice sites, a 
very appealing feature which would lead to a very efficient quantum algorithm once
a way is found to boost the single-step success probability and bring it close to the unit value.

A natural candidate to this end is the oblivious amplitude amplification 
algorithm~\cite{berry_simulating_2015}, which could in principle pull up the success 
probability close to the unit value. The challenge is then to develop a version
of such an algorithm capable of handling non-unitary matrices.
As an alterantive, one could borrow ideas from the theory of open quantum systems, i.e.
augment the LB equation with unitarity-restoring extra degrees of freedom. 
For instance one could add an extra population, $g_i$ (mirror reservoir), obeying the same LB equation
but with a collision term of opposite sign, so that the dynamics of the sum of the 
two, $h_i=f_i+g_i$, is collision free, hence unitary. 
One would then apply oblivious amplitude amplification to both $f_i$ and $g_i$ and focus
the attention on $f_i$ only.    
A second  possibility is to expand the CLB matrix onto a second-quantized basis of annihilation and
generation operators and apply block-encoding to each term in the expansion separately. 
The prospective advantage is twofold; first, the expansion is likely to converge significantly faster
than the Pauli basis expansion and second, each term is a one-sparse matrix, hence 
it can be implemented by the same block-encoding technique with 
a $O(1)$ success probability, since each matrix would require a single ancilla qubit. 
The challenge is to collect the cumulative effect of the various terms without losing efficiency.   

A third (very optimistic) option is to analyse the nature of the errors introduced by the 
dissipative update, in the hope that they may display some sort 
of fault-tolerance or maybe even self-healing properties.  
This appears very unlikely, but a close investigation
of the concrete effects of the failed dissipative updates on the dynamics of the quantum
system may reveal unexpected features and possibly stimulate
new ideas to increase the success probability.

All options above make an interesting topic for future work.

\section*{Acknowledgements}

We acknowledge financial support form the Italian National
Centre for HPC, Big Data and Quantum Computing (CN00000013) and from the NSF STAQ project under award NSF PHY-232580. 
One of the authors (SS) is grateful to the Physics and Astronomy Department of Tufts University
for kind hospitality and financial support.

\bibliographystyle{unsrt}
\bibliography{bibliography}

\end{document}